\documentstyle[12pt,amsfonts]{article}
\newcommand{\mbf}[1]{\mbox{\boldmath$ #1$}}

\newcommand{\be}{\begin{equation}}
\newcommand{\ee}{\end{equation}}
\newcommand{\ba}{\begin{eqnarray}}
\newcommand{\ea}{\end{eqnarray}}

\begin{document}

\begin{center}

{\Large\bf Wave packet propagation by the Faber polynomial
  approximation in electrodynamics of passive media}

\vskip 1cm
{\Large Andrei G. Borisov ${}^{a,}
$\footnote{email: {\sf borisov@lcam.u-psud.fr}}
and  Sergei V. Shabanov ${}^{b,}
$\footnote{email: {\sf shabanov@phys.ufl.edu}}}

\vskip 1cm

${}^a${\it Laboratoire des Collisions Atomiques et Mol\'eculaires,
UMR CNRS-Universit\'{e} Paris-Sud 8625, B\^{a}t. 351,
Universit\'{e} Paris-Sud, 91405 Orsay CEDEX, France  }
\vskip 0.3cm

${}^b$
{\it Department of Mathematics,
University of Florida, Gainesville, FL 32611, USA}

\end{center}

\begin{abstract}
Maxwell's equations for propagation of electromagnetic waves in
dispersive and absorptive (passive) media are represented in the
form of the Schr\"odinger equation $i\partial \Psi/\partial t = {
H}\Psi$, where ${ H}$ is a linear differential operator
(Hamiltonian) acting on  a  multi-dimensional vector $\Psi$ composed of
the electromagnetic fields and auxiliary matter fields describing
the medium response. In this representation, the initial value
problem is solved by applying the fundamental solution $\exp(-itH)$
to the initial field configuration. The Faber polynomial
approximation of the fundamental solution is used to develop a
numerical algorithm for propagation of broad band wave packets in
passive media. The action of the Hamiltonian on the wave function
$\Psi$ is approximated by the Fourier grid pseudospectral method.
The algorithm is global in time, meaning that the
entire propagation can be carried out in just a few time steps. A typical
time step is much larger than that in finite differencing schemes,
$\Delta t_F \gg \|H\|^{-1}$. The accuracy and stability of the
algorithm is analyzed. The Faber propagation method is
compared with the Lanczos-Arnoldi propagation method with an example
of scattering of broad band laser pulses on a periodic grating made
of a dielectric whose dispersive properties are described by the
Rocard-Powels-Debye model. The Faber algorithm is shown to be more
efficient. The Courant limit for time stepping, $\Delta t_C \sim
\|H\|^{-1}$, is exceeded at least in 3000 times in the Faber
propagation scheme.

\end{abstract}

\newpage

\section{Introduction}

Many time-domain algorithms for numerical simulations of the
broad band wave packet propagation in electrodynamics of passive
media and/or quantum mechanics use a time stepping, that is, given a
configuration of the system at time $t$, a time-domain algorithm
produces the system configuration at time $t+\Delta t$, where the
time step $\Delta t$ is determined by conditions resulting from the
algorithm stability and required accuracy. For instance, in a finite
differencing approach, such as, e.g., the classical leapfrog scheme,
the time step is bounded from above by the stability condition (the
Courant limit), $\Delta t \leq \Delta t_C$. The upper bound $\Delta
t_C$ is typically determined by the time a signal needs to propagate
through an elementary cell of the spatial grid, which is 
by several orders of magnitude smaller than the total propagation
time \cite{taflove}. There is a class of problems in numerical
electromagnetism where the wave packet dynamics at intermediate
times is not of significant interest, but rather the final state is
important. Computing the scattering matrix would give one such
example. A related and more sophisticated example would be
simulations of the  broad band wave packet propagation in random
media \cite{rand}. To obtain a numerical solution of  the initial-value problem
in this case, the propagation must be carried out 
multiple times for every
(random) state of the medium in order to perform the statistical
averaging over the medium states. Clearly, a global time-domain
algorithm ($\Delta t \gg \Delta t_C$) would be of great help in
reducing computational costs.

The present work offers a global time-domain algorithm for solving
initial value problems for Maxwell's equations for passive media
whose dispersive and absorptive properties can be described by
suitable Lorentz, or Rocard-Powels-Debye, or Drude models. The basic
idea of our approach can be summarized as follows. In Section 2, the
Maxwell equations are cast in the form of the Schr\"odinger equation
\be 
\label{1.1} 
i\,\frac{\partial \Psi}{\partial t} = H\Psi\ , 
\ee
where $\Psi$ is a multidimensional vector field whose components are
electromagnetic fields and a set of auxiliary fields that describe
the medium response to applied electromagnetic fields (e.g., the
medium polarization), and $H$ is a linear differential
operator that depends on
the medium dispersive and absorptive properties. Its spectrum is
real if no attenuation is present, and has a negative imaginary part
otherwise. The squared $L_2$ norm of $\Psi$ is proportional to the
electromagnetic energy of the wave packet.

If $\Psi_0$ is the initial wave packet configuration, then $\Psi(t)$
can be found by using the fundamental solution of Eq. (\ref{1.1})
\be 
\label{1.2} 
\Psi(t) = e^{-itH}\Psi_0\ . 
\ee 
Given some (grid)
approximation of the spatial dependence of $H$ and $\Psi$, Eq.
(\ref{1.2}) provides a numerical solution of the initial value
problem. In what follows the same letters are used for spatial
continuum and grid representations of the Hamiltonian and wave
functions, unless noted otherwise. An exact solution of the initial
value problem is understood here in the sense of (\ref{1.2}) where
$H$ is a finite matrix obtained from the continuous Hamiltonian by
means of a suitable, sufficiently accurate, spatial (grid)
representation.

If $H$ can be directly diagonalized, then (\ref{1.2}) gives an exact
solution for any value of $t >0$. But this is precisely what one
wants to avoid in numerical simulations because the matrix $H$ is
typically huge and the direct diagonalization is too expensive, if
impossible at all. For this reason, time domain algorithms use the
semigroup property of the fundamental solution: $\exp(-itH)=
[\exp(-i\Delta tH)]^N$, where $\Delta t=t/N$ with an integer $N$
being the number of time steps. For a sufficiently small time step
$\Delta t$, typically, $\Delta t \sim \|H\|^{-1}$, where $\|H\|$ is
the (matrix) norm of $H$, the action of the infinitesimal evolution
operator $\exp(-i\Delta tH)$ on the state vector $\Psi$ can be
approximated by various means that do not require any direct
diagonalization of $H$.

Section 3 is devoted to an algorithm that  involves neither a direct
diagonalization of $H$ nor many time steps. It is based on the well
known approximation of an analytical function by the Faber
polynomial series \cite{faber} (see also the textbooks \cite{mark}).
The Faber approximation method has been applied to quantum
scattering problems \cite{hoffman} to compute the causal Green's
function for the Schr\"odinger equation. 
The Faber polynomial approximation of the exponential of a
non-Hermitian operator has also been used to solve the initial value
problem for the Liouville - von Neumann equation that describes the
time evolution of the density matrix in statistical systems
\cite{kosloff2a,kosloff2b}.
In the case when the
spectrum of $H$ is real, the approximation yields the well known
Chebyshev propagation method that has been developed to study wave
packet dynamics in quantum systems \cite{kosloff1,Review,Ps} and
later used in electrodynamics of  non-dispersive media \cite{cole}.

We apply the Faber propagation scheme to solve initial value
problems in electrodynamics of passive media reformulated in the
form of the Schr\"odinger equation (\ref{1.1}) with a non-Hermitian
Hamiltonian, 
\be 
\label{1.3} 
\Psi(t+\Delta t_F) = e^{-i\Delta
t_FH}\Psi(t) \approx \sum_{k=0}^{n} c_k(\Delta t_F) F_k(H)\Psi(t)\ .
\ee 
Here $c_k(\Delta t_F)$ are the expansion coefficients and
$F_k(H)$ are Faber polynomials. The action of $F_k(H)$ on $\Psi(t)$
can be computed recursively. The recursion relation depends on the
choice of the family of Faber polynomials. The latter, in turn, is
motivated by spectral properties of $H$. An important point to note
is that the expansion (\ref{1.3}) gives an accurate approximation
for the fundamental solution for large values of $\Delta t_F\,
>\!\!>\,\Delta t_C\sim \| H\|^{-1}$ and, hence, the propagation can
be done in just a few time steps. The Faber series (\ref{1.3}) is
known to converge exponentially as the approximation order $n$
increases. The accuracy of the algorithm is assessed in Section 4.
In Sections 5 and 6 the algorithm is applied to scattering of broad
band laser pulses on a dielectric grating. Dispersive properties
of the grating material are described by the Rocard-Powels-Debye
model with a single pole. The frequency band of the initial pulse is
chosen to cover the anomalous dispersion range (the pole) of the
dielectric. The Faber propagation scheme is shown to be more
efficient than the Lanczos-Arnoldi propagation scheme applied
earlier to the same system \cite{as2}. The Courant
limit can be exceeded in at least 3000 times, $\Delta t_F \geq
3000\Delta t_C$. Due to the exponential convergence of the algorithm
it can be used as a benchmark for testing various time propagation
schemes. Note also that it can be applied with any suitable
finite-dimensional approximation of the Hamiltonian $H$ (finite
elements, or finite differencing, or any spectral representation).
In our simulations, the Fourier grid pseudospectral representation
of $H$ has been used \cite{boyd,kosloff}.

\section{Maxwell equations in the Hamiltonian form}

\setcounter{equation}0 Let ${\bf D}$ and ${\bf B}$ be electric and
magnetic inductions, respectively, and ${\bf E}$ and ${\bf H} $ the
corresponding fields. When no external currents and charges are
present, the dynamical Maxwell's equations read \be \label{2.1}
\dot{\bf D} = c \mbf{\nabla}\times {\bf H}\ ,\ \ \ \dot{\bf B} = -c
\mbf{\nabla}\times {\bf E}\ . \ee The over-dot denotes the partial
derivative with respect to time, and $c$ is the speed of light in
the vacuum. Equations (\ref{2.1}) have to be supplemented by the
Gauss law $\mbf{\nabla}\cdot{\bf  D}=0$ and also by
$\mbf{\nabla}\cdot{\bf B}=0$. Relations between the fields and
inductions are determined by physical properties of the medium in
question.

As an example we consider the Rocard-Powles-Debye model dielectric
(the ionic crystal model \cite{Ionic1,Ionic2}) with one resonance,
which is used in our numerical simulations. The case with multiple
resonances can be studied in a similar fashion. In this model ${\bf
H}={\bf B}$, and the Fourier harmonics of the electric field and
induction of frequency $\omega$ are related by ${\bf D}(\omega) =
\varepsilon(\omega) {\bf E}(\omega)$ where the dielectric constant
is given by \be \label{2.2} \varepsilon (\omega )=\varepsilon
_{\infty }+\frac{(\varepsilon _{0}-\varepsilon _{\infty })\omega
_{T}^{2}}{\omega _{T}^{2}-\omega ^{2}-i\eta \omega }\ , \ee with
$\varepsilon _{\infty ,0}$ being constants, $\omega _{T}$ the
resonant frequency, and $\eta $  the attenuation. Let ${\bf P}$ be
the dispersive part of the total polarization vector of the medium.
Then ${\bf D}=\varepsilon _{\infty }{\bf E}+{\bf P}$. By using the
Fourier transform, it is straightforward to deduce that ${\bf P}$
satisfies the second-order differential equation
\begin{equation}
\label{2.3}
\ddot{{\bf P}}+\eta \dot{{\bf P}}+\omega _{T}^{2}{\bf P}=\varepsilon
_{\infty }\omega _{p}^{2}{\bf E}\ ,
\end{equation}
where $\omega _{p}^{2}=(\varepsilon
_{0}-\varepsilon _{\infty })\omega _{T}^{2}/\varepsilon _{\infty }$ if $%
\varepsilon _{0}-\varepsilon _{\infty }$ is positive, otherwise, $\omega
_{p}^{2}\rightarrow -\omega _{p}^{2}$ in (\ref{2.3}). Equation (\ref{2.3}) must
be solved with  zero initial conditions, ${\bf P}=\dot{{\bf P}}=0$ at $%
t=0 $.

Define a set of auxiliary fields ${\bf Q}_{1,2}$ by ${\bf P}=\sqrt{%
\varepsilon _{\infty }}\omega _{p}{\bf Q}_{1}/\omega _{T}$ and $\dot{{\bf Q}}%
_{1}=\omega _{T}{\bf Q}_{2}$.
Maxwell's equations and (\ref{2.3}) can be written as the
Schr\"{o}dinger equation (\ref{1.1}) in which the wave function and the
Hamiltonian are defined by
\begin{equation}
\label{2.4}
\Psi =\pmatrix{ \varepsilon_\infty^{1/2}{\bf E}\cr {\bf B}\cr {\bf Q}_1\cr
{\bf Q}_2}\ ,\ \ \ { H}=\pmatrix{0& ic
\varepsilon_\infty^{-1/2}\mbf{\nabla}\times &0& -i\omega_p\cr
-ic\mbf{\nabla}\times \varepsilon_\infty^{-1/2}&0&0&0\cr 0&0&0&i\omega_T\cr
i\omega_p&0& -i\omega_T&-i\eta }\ .
\end{equation}
Here $\varepsilon _{\infty ,0}$ are set to one in the vacuum, and to
some specific values in the medium in question. The squared $L_2$
norm of the wave function is proportional to the total
electromagnetic energy of the wave packet. When  attenuation is not
present, $\eta =0$, the Hamiltonian is Hermitian relative to the
conventional $L_2$ scalar product, and the norm (or energy) is
conserved.

In our simulations, an absorbing layer of a conducting medium has
been introduced  at the grid boundaries to prevent reflections of the
wave packet. The conductivity $\sigma$ of the layer depends on
position. The induced current in a conducting media has the form
$\sigma{\bf E}$. Hence, in the presence of the conducting layer the
Hamiltonian (\ref{2.4}) is modified by inserting $-4\pi
i\sqrt{\varepsilon_{\infty}}\sigma$ in place of zero in the
upper-right corner. Further details can be found in our earlier
works \cite{svs,as1,as2}.

\section{Faber polynomial propagation scheme}
\setcounter{equation}0

Let $D$ be a bounded, closed continuum in the complex
plane such that the complement of $D$ is simply connected
in the extended complex plane and contains the point
at $z=\infty$ (e.g., a polygon, an ellipse, etc.).
By the Riemann mapping theorem \cite{mark}, there exists a conformal
mapping $\xi$ which maps the complement of a closed disk
with center at the origin and radius $\rho$ onto the
complement of $D$, satisfying the normalization condition,
$\xi(w)/w \rightarrow 1$ as $|w|\rightarrow \infty$.
Then its Laurent expansion at $\infty$ is given by
\be
\label{3.1}
\xi(w) = w + \sum_{k\geq 0} \gamma_kw^{-k}\ .
\ee
The radius $\rho$ of the disk is called
the logarithmic capacity of $D$. This quantity plays an
important role in the accuracy analysis given below.
The family of Faber polynomials $F_k$ associated with a
conformal mapping $\xi$ is defined via the recursion
relation
\be
\label{3.2}
F_{k+1}(z) = zF_k(z) - \sum_{j=0}^k \gamma_jF_{k-j}(z) -k\gamma_k\ ,
\ \ \ \ F_0(z)=1\ .
\ee
For a function $f(z)$ that is analytic at every point of $D$,
the Faber series 
$$
f(z)=\sum_{k=0}^\infty c_kF_k(z) 
$$ 
is defined by
\be
\label{3.3}
c_k = \frac{1}{2\pi i}\int_{|w|=R} \
\frac{f(\xi(w))}{w^{k+1}}\ dw\ ,
\ee
where $R>\rho$ is sufficiently small that $f$ can be extended
analytically to the contour $\Gamma_R$ being the image
of the circle $|w|=R$ under the conformal mapping $\xi$.
The value $R=\rho$ is acceptable if $\xi$ can be extended
continuously to the circle $|w|=\rho$ (e.g., when
the boundary of $D$ is a closed simple curve with no
self-intersections (a Jourdan curve)).
The Faber series converges uniformly and absolutely
to $f$ on every region bounded by $\Gamma_R$ to which
$f$ can be extended analytically \cite{covari}.
This theorem establishes mathematical foundations for
the Faber polynomial approximation (\ref{1.3}) of the
fundamental solution of (\ref{1.1}).

The Faber polynomial algorithm for
solving initial value problems for (\ref{1.1})
is as follows.
First, choose a (Jourdan) contour $\Gamma$  that encloses
the spectrum of $H$. Some criteria for choosing a contour
are discussed in the next section.
Second, find the corresponding conformal mapping $\xi$.
In particular, if $\Gamma$ is a polygon, this task
can be accomplished by the Schwartz-Christoffel transformation.
For complicated polygons, there is a numerical algorithm
to do so \cite{trefethen}.
Next, the Faber
expansion coefficients $c_k(\Delta t_F)$ are computed
by means of (\ref{3.3}) where $f(z) = \exp(-i\Delta t_Fz)$.
The action of the Faber polynomials of $H$ on $\Psi(t)$
in (\ref{1.3})
is computed using the recursion relation (\ref{3.2}).
Let $\Phi_k = F_k(H)\Psi(t)$. Then
\be
\label{3.5}
\Phi_{k+1} = (H-k\gamma_k)\Phi_k -
\sum_{j=0}^k\gamma_j \Phi_{k-j}\ ,
\ee
where $\Phi_0 = \Psi(t)$ and $\Phi_1 = (H-\gamma_0)\Phi_0$.
The series (\ref{1.3}) converges uniformly on the entire
spectral range of $H$.
In order to make the algorithm memory friendly, it is
desired to make the sequence of $\gamma_k$ not only
finite, but also as short as possible. In Section 5
we apply this algorithm to the Hamiltonian (\ref{2.4})
and choose an ellipse to enclose its spectrum.

From the numerical point of view, the recursion
relation (\ref{3.5}) is, in general, unstable because the minimax norm
of Faber polynomials grows rapidly as their order
increases,
$\max_D |F_k(z)| \leq 2\rho^k$ (see \cite{starke}).
In other words, the norm of $\Phi_k$
would grow exponentially, while the decay of $c_k(\Delta t_F)$
still provides the convergence of (\ref{1.3}). However,
in a numerical implementation of (\ref{3.5}), one might 
encounter floating point exceptions with a subsequent 
loss of accuracy.  To avoid this instability,
the Hamiltonian $H$ must be scaled so that its spectrum
lies in the domain whose logarithmic capacity is one.
If $\beta$ is the scaling factor, then
$\exp(-i\Delta t_FH) = \exp(-i\Delta t_sH_s)$
where $H_s = H/\beta$
and $\Delta t_s = \beta\Delta t_F$. Thus, in the recursion relation
(\ref{3.5}) the scaled Hamiltonian $H_s$ 
and the sequence $\gamma_k$ generated by the conformal mapping
(\ref{3.1}) with $\rho =1$
must be used,
while the expansion coefficients in the Faber series
(\ref{1.3}) are determined by
\be
\label{ck}
c_k(\Delta t_s) =
\frac{1}{2\pi}\int_{0}^{2\pi}
\exp\left[-i\Delta t_s\xi(e^{i\varphi})\right]\, e^{-ik\varphi}
d\varphi
\ .
\ee
Note that the exponential $f(z) = \exp(-i\Delta t_sz)$
is an analytic function in the entire complex plane 
so that, assuming $\Gamma$ to be a Jourdan curve, one can set
$R=\rho$ in (\ref{3.3}) and use the fact that the spectrum
of the scaled Hamiltonian lies in a domain with $\rho =1$
and therefore $w=e^{i\varphi}$ in (\ref{3.3}). Clearly,
the scaling factor $\beta$ must chosen as small as
possible to allow for larger time steps
$\Delta t_F = \Delta t_s/\beta$.

\section{Accuracy and efficiency assessment}
\setcounter{equation}0

The range $R_H$ of $H$ is
a set of complex numbers $(\Psi,H\Psi)/\|\Psi\|^2$ obtained
for all normalizable wave functions $\Psi$.
Here $(\cdot,\cdot)$ denotes a
scalar product, and $\|\cdot\|$ is the norm associated
with it.
The norm of the  resolvent of $H$ is bounded by \cite{resolvent}
\be
\label{4.7}
\|(z-H)^{-1}\| \leq [d(z,R_H)]^{-1}\ ,
\ee
where $d(z,z')=|z-z'|$ is the distance on the complex
plane, and the distance between $z$ and a set $R_H$ is
defined as $\min_{z'\in R_H}d(z,z')$.
Let $\Gamma$ be any closed (Jourdan) curve enclosing
the spectrum of $H$. Let $P_n$ be a polynomial
of order $n$ that is used to approximate
the fundamental solution of (\ref{1.1}),
that is, $\exp(-itH)\Psi_0\approx P_n(H)\Psi_0$.
By making use of the Cauchy theorem, it
is straightforward to see that the accuracy
of the approximation is bounded by
\ba
\nonumber
\|e^{-itH}\Psi_0 - P_n(H)\Psi_0\| &=&
\left\| \frac{1}{2\pi i}\int_\Gamma
\frac{e^{-itz}-P_n(z)}{z-H}\ \Psi_0\, dz\right\| \\
\label{3.4}
&\leq& C_\Gamma \|\Psi_0\|
\max_{z\in \Gamma}\left| e^{-itz}-P_n(z)\right|
\equiv \epsilon_n(\Gamma)\|\Psi_0\|\ ,
\ea
where the constant $
C_\Gamma = {L_\Gamma}/[2\pi d(\Gamma,R_H)]$ and
$L_\Gamma$ is the length of $\Gamma$.
To find $C_\Gamma$, Eq. (\ref{4.7}) has been used.
Note that
$C_\Gamma$ depends on $H$ and $\Gamma$, but is
independent of the approximation
order $n$. Hence, it follows from (\ref{3.4}) that
the error of the
polynomial approximation of the solution of the
initial value problem for (\ref{1.1}) can be made
as small as desired because
the Faber polynomial approximation $P_n(z)$ converges to
$\exp(-itz)$ absolutely and uniformly in D.

In addition, it is worth noting that
the Faber polynomial approximation provides
the so called ``near best'' polynomial approximation
of an analytic function.
By definition, $\|f\|_\infty = \max_{z\in D}|f(z)|$.
The maximum principle for analytic functions states \cite{mark} that
if $f$ is analytic in $D$ and continuous in the closure
of $D$, then $|f|$ cannot attain its maximum at interior
points of $D$.
According to (\ref{3.4}) and  the maximum principle for functions
analytic in $D$ bounded by $\Gamma$,
the accuracy $\epsilon_n(\Gamma)$
 of a polynomial approximation
of an analytic function $f$ of a matrix $H$ (in our case, $f(z)=
\exp(-itz)$) is
\be
\label{4.1}
\epsilon_n(\Gamma) = C_\Gamma \|e^{-itz}-P_n(z)\|_\infty\ .
\ee
The fundamental theorem for polynomial approximations of
functions analytic in the interior of $D$ and continuous in $D$
states that there exists a unique best minimax polynomial approximation
$P_n^f$ to $f$, that is, \cite{ellacott}
\be
\label{4.2}
\|f-P_n^f\|_\infty \leq \|f-P_n\|_\infty
\ee
for any polynomial $P_n$ of order $n$. In practice, it is not easy
to find $P_n^f$. Suppose we choose some polynomial approximation, that
is, we define a projection operator ${\cal P}_nf = P_n$.
In particular, for Faber
polynomials ${\cal P}_n={\cal P}_n^F$, and
${\cal P}_n^Ff$ is given by the truncated Faber series.
Then it
follows from the identity $f-{\cal P}_nf= f-P_n^f + {\cal
  P}_n(P_n^f-f)$ that
\be
\label{4.3}
\|f-{\cal P}_nf\|_\infty \leq (1+\|{\cal P}_n\|)\|f-P_n^f\|_\infty\ .
\ee
Thus, our polynomial approximation appears to be ``near best'',
provided the norm of the projection 
operator ${\cal P}_n$
is not so large. For Faber polynomials,
one can show that \cite{ellacott,mason}
\be
\label{4.4}
\|{\cal P}_n^F\| \leq \frac{V}{\pi}\left(\frac{4}{\pi^2}\,\ln n
+B\right)
\ee
for $n\geq 1$. Here $B \approx 1.773$ and
$V=\int_\Gamma |d\theta(z)| \geq 2\pi$
and $\theta(z)$ is the angle that is made by a line tangent
to $\Gamma$ with the positive real axis. For a convex $D$,
$V=2\pi$ by the Radon theorem. In our simulations, $D$ is
an ellipse, which is convex, therefore
\be
\label{4.5}
\|{\cal P}_n^F\| < 9\ ,\ \ \ n\leq 835\ .
\ee
Equation (\ref{4.5}) shows that by using the Faber
polynomial approximation to $f$ we do not loose more than
one decimal place in accuracy as compared with the
best minimax polynomial approximation.
In this case, one can also show that \cite{mark}
\be
\label{4.6}
\|f-{\cal P}_n^Ff\|_\infty \leq
\frac {(\rho/R)^{n+1}V}{\pi(1-\rho/R)}\,
\max_{z\in \Gamma_R}|f(z)|\ ,
\ee
for any domain bounded by $\Gamma_R$, $R>\rho$, to which
$f$ can be extended analytically.
Thus, the Faber series (\ref{1.3}) converges exponentially as
the approximation order $n$ increases.
From (\ref{4.6}) some basic principles
for choosing the contour $\Gamma$ follow.

First, because of the exponential convergence of the  Faber series,
it is desired to make
the logarithmic capacity $\rho$ as small as
possible. Alternatively, if $\rho$ is set to one,
the scaling factor $\beta$ must be as small as
possible, that is, the contour should enclose the spectrum
of $H$ as tight as possible. In principle, if the structure
of the spectrum of $H$ (or, at least, its range) is roughly
known, one can find a polygon that tightly encloses the
spectrum. The corresponding conformal mapping can be
computed numerically \cite{trefethen}. The unfortunate
feature of this approach is that the infinite Laurent
series (\ref{3.1}) is required. Hence, the recursion
relation (\ref{3.5}) becomes memory unfriendly in numerical
simulations: All the preceding $\Phi_k$ must be kept in
the operational memory. Thus, when choosing the contour,
one should compromise between
the approximation order and the memory use efficiency of
the algorithm \cite{hoffman}.

Second, if possible,
the contour $\Gamma$ should not go too far into the upper part
of the complex plane to avoid the exponential growth
of the factor $\max_\Gamma |\exp(-i\Delta t_sz)|$ and to allow
for larger time steps. 
Note that the necessary accuracy
can still be reached, even if the contour goes through the
upper part of the complex plane, by increasing the approximation
order $n$. The latter, however, would lead to a less efficient
propagation scheme because more operations per time step is
required.

\section{The case of an elliptic contour}
\setcounter{equation}0

Faber polynomials associated with an elliptic contour
have the simplest (shortest) recursion relation \cite{mark}.
For this reason this family of the Faber polynomials have
been used in many aforementioned applications in quantum 
and statistical mechanics.
Here we use the Faber polynomials associated with an ellipse
to illustrate the Faber propagation scheme in electrodynamics of passive
media.

Consider
$\Gamma$ being an ellipse $(x-x_0)^2/a^2 + (y-y_0)^2/b^2 =1 $
where $z=x+iy$. The ellipse is an image of the circle $|w|=\rho$
under the conformal mapping
\be
\label{3.6}
\xi(w) = w + \gamma_0 + \gamma_1/w\ ,
\ee
where $a= \rho +\gamma_1/\rho$,
$b= \rho -\gamma_1/\rho$, and $\gamma_0 = x_0 +iy_0$
is the center of the ellipse. The logarithmic
capacity of an ellipse is $\rho = (a+b)/2$ and
$\gamma_1 = \rho (a-b)/2$.
We choose $\rho=1$ so that
$$
\gamma_1 = 1-b\ .
$$
In this case,
the optimization
parameters are the scaling factor $\beta$
and the number $b+y_0$ that determines the factor
$\max_{\Gamma}|\exp(-i\Delta t_sz)|=\exp[\Delta t_s(b+y_0)]$ in the accuracy
(\ref{4.6})
of the Faber approximation.

The
recursion relation (\ref{3.5}) associated with the elliptic contour
has only two terms
\be
\label{3.7}
\Phi_{k+1} = (H_s-\gamma_0)\Phi_k - \gamma_1\Phi_{k-1} \ ,\ \ \
k>1\ ,
\ee
where $\Phi_0=\Psi(t)$ and $\Phi_1 = (H_s-\gamma_0)\Phi_0$.
The Faber expansion coefficients have the form
\be
\label{3.8}
c_k(t_s) = \left(\frac{-i}{\sqrt{\gamma_1}}\right)^k
e^{-i\Delta t_s\gamma_0} J_k(2t_s\sqrt{\gamma_1})\ ,
\ee
Here $J_k$ is the Bessel function. When computing the
integral (\ref{ck}) we assume that $\gamma_1>0$ (which
is consistent with the spectral properties of 
the Hamiltonian $H$ used in our simulations).
The exponential convergence
of the Faber series can easily be seen  from the exponential
decay of the Bessel function for $k> 2\Delta t_s\sqrt{\gamma_1}$.

The Hamiltonian (\ref{2.4}) cannot have eigenvalues with positive imaginary
parts, otherwise the energy of the wave packet
(the squared norm of $\Psi$) would
increase with time, which is not
possible in passive media. Hence, by physical reasons, the spectrum
of the Hamiltonian lies in the lower half of the complex plane.
It is also clear that the spectrum of the
Hamiltonian is symmetric about the imaginary axis (for every
direction in space, there are incoming and outgoing waves). Hence,
we set $x_0=0$. The spectrum of $H$ lies in a rectangle
$[-E_m,E_m]\times [-v,0]$ with $E_m$ and $v$ to be determined
below. Our strategy is to find an ``optimal'' ellipse with $\rho =1$
that contains a scaled rectangle $[-E_s,E_s]\times [-v_s,0]$, where
$E_s = E_m/\beta$ and $v_s=v/\beta$.

First, we determine the bounds, $E_m$ and $v$, on the spectral
range of $H$.
Let
$z_\psi\ = (\Psi,H\Psi)/\|\Psi\|^2$ be a point
in $R_H$. Let $H=H_0 -iV$ where $H_0=(H+H^\dagger)/2
=H_0^\dagger$ and $V=i(H-H^\dagger)/2 =V^\dagger$
is positive semidefinite. Then
\be
\label{5.1}
E_m= \max_{\Psi} {\rm Re}\, z_\psi =
\max_{\Psi}\ (\Psi,H_0\Psi)/\|\Psi\|^2 =
\|H_0\|\ .
\ee
Thus, $E_m$ is the maximal eigenvalue of
$H_0$ because $H_0$ is Hermitian. It can
be found by the standard numerical procedure.
If $\Psi_n = H_0\Psi_{n-1}$ for $n=1,2,...$ for
some initial vector $\Psi_0$, the sequence
$\|\Psi_{n}\|/\|\Psi_{n-1}\|$
converges to the maximal eigenvalue $E_m$
of $H_0$ as $n$ increases. 
A rough estimate for $E_m$ can also be obtained
by noting that the maximal wave vector supported
by the grid in the Fourier pseudospectral
representation is $k_{max} = \pi/a_{min}$ with $a_{min}$
being the smallest grid step (if a non-uniform grid
is used). Hence, $E_m \approx ck_{max}$.
Similarly,
\be
\label{5.2}
v=\max_\Psi (-{\rm Im}\, z_\psi) =
\max_\Psi\   \ (\Psi,V\Psi)/\|\Psi\|^2 =
\max\{\ 4\pi\sqrt{\varepsilon_\infty}\sigma_{max},\ \eta\ \} =
4\pi\sqrt{\varepsilon_\infty}\sigma_{max}\ ,
\ee
where $\sigma_{max}$ is the maximal value
of conductivity of the absorbing layer.
Here we have used that fact that $V$ is
diagonal and the medium attenuation $\eta$
is small compared to $\sigma_{max}$.

By the symmetry, the center of the ellipse is set to coincide with the center
of the rectangle,
$$
\gamma_0 = -iv_s/2\ .
$$
An ellipse that contains the rectangle vertices
should
satisfy the following condition
\be
\label{ell}
\frac ab = \frac{E_s}{\sqrt{b^2 - |\gamma_0|^2}}\ .
\ee
Since $\rho = 1$, $a=b-2$. Equation (\ref{ell})
relates $b$ and the scaling factor $\beta$.

As has been argued above, to increase the
time step $\Delta t_F =\Delta t_s/\beta$, the
scaling factor $\beta$ must be minimal.
So, one can take $b$ for which $\beta$
attains its minimal value.
The smallest $\beta$ is reached
when
\be
\label{opt}
\frac ba = \left(\frac{|\gamma_0|}{E_s}\right)^{2/3}=
\left(\frac{v}{2E_m}\right)^{2/3}\ .
\ee
Observe that if $v=0$, that is, if the spectrum of $H$
is real, the optimal ellipse has $b=0$ and $\beta =
E_m$. In this case, the Faber polynomial series is nothing
but the Chebyshev polynomial series.
Unfortunately, when $v\neq 0$ by making $\beta$ smaller
we increase the number $b+y_0$, that is,
the ellipse gets farther into the upper half
of the complex plane and higher orders of
the Faber approximation are needed to achieve
desired accuracy according to (\ref{4.6}).
So, in our simulations we take $\beta$ larger
than its minimal value and thereby
reduce $b+y_0$ by making $b$ smaller
(see next Section for details).

\section{Applications to nanostructured periodic materials}

\setcounter{equation}0

As an example of possible applications of the present method to
photonics, the Faber propagation scheme associated with an elliptic
contour is applied to scattering of broad band wave packets on
nanostructured periodic materials, the subject of current interest
in photonics
\cite{Photonics}. We consider a grating made of a periodic array of
ionic crystal cylinders in vacuum. This system has been previously
studied by the Lanczos-Arnoldi time propagation scheme \cite{as2}.
In particular, the role of trapped modes (guided wave resonances)
and polaritonic excitations in transmission and reflection
properties of the grating in the infrared range has been elucidated.
Apart from illustrating the Faber propagation scheme, our primary
interest is to compare its efficiency with the efficiency of the
Lanczos-Arnoldi propagation scheme.

 The geometry of the system is
sketched in the inset of Fig. 2.
The system has a translation symmetry along one of the Euclidean
axes, chosen to be the $y$ axis. It is periodic along the $x$ axis
with period $D_{g}$, while the $z$ direction is transverse to the
grating. The packing density $R/D_{g}=0.1$, where $R$ is the radius
of cylinders and $D_{g}=10.8~\mu m$ is the grating period. The
broad band wave packet is represented by a Gaussian
pulse that is about $38$ fs long and has 
the carrier frequency of $173$ meV.
It propagates along the $z$ axis and is linearly
polarized with the electric field oriented along the $y$ axis, i.e.,
parallel to the cylinders (the so called TE polarization). The
spectrum of the wave packet is concentrated in a wavelength domain
$\lambda \geq D_{g}$ such that the scattering is dominated by the
zero diffraction mode (the reflected and transmitted beams propagate
mainly along the $z$-axis). A change of variables is used in both
$x$ $(x=f_{1}(x_1 ))$ and $z$ $(z=f_{2}(x_2 ))$ coordinates to
enhance the sampling efficiency in the vicinity of medium interfaces
so that the boundary conditions are accurately reproduced by the
Fourier grid pseudospectral method. A typical size of the mesh
corresponds to $-17.3D_{g}\leq z\leq 15.3D_{g}$, and $-0.5D_{g}\leq
x\leq 0.5D_{g}$ with, respectively, $384$ and $64$ mesh points. Note
that, because of the variable change, a uniform mesh in the
auxiliary coordinates $(x_1 ,x_2 )$ corresponds to a non-uniform
mesh in the physical $(x,z)$ space. The Lanczos-Arnoldi time
propagation is carried out with a fixed time step $\Delta
t_{L}=0.138$ fs. The propagation by 
the Faber method has been done with different time steps $\Delta
t_{F}=j\Delta t_{L}$, with $j=25,50,100,200,400,$ and $1000$ (see
below).

The dielectric function of the ionic crystal material is
approximated by the single oscillator model (\ref{2.3}). Following
the work \cite{Ionic1}, we chose the parameters representative for
the beryllium oxide: $\varepsilon _{\infty }=2.99$, $\varepsilon
_{0}=6.6$, $\omega _{T}=87.0$ meV, and the damping $\eta =11.51$
meV. Thus, for $D_{g}=10.8\mu m$ two types of resonances can be
excited in the system within the frequency domain covered by the
incident pulse. Structure resonances are characteristic for periodic
dielectric gratings. They are associated with the existence of
guided wave modes \cite{Wood,DielGratings}. As has been demonstrated
previously, in the absence of losses, structure resonances lead to
100\% reflection within a narrow frequency interval(s) for
wavelengths $\lambda \sim D_{g}$. The second type of resonances
arise because of polaritonic excitations for wavelengths $\lambda
\sim D_{T}=2\pi c/\omega _{T}=26.9~\mu m$. These are associated with
substantial energy losses in the ionic crystal material. A detailed
discussion of the transmission and reflection properties of this
grating can be found in \cite{as2}.

Figure 1a shows the elliptic contour used in our simulations. Its
logarithmic capacity is one, $\rho =1$, and the corresponding
conformal mapping (\ref{3.6}) reads $\xi(w)=w-0.005~i+0.99/w$ so
that $b=0.01$. The scaling factor $\beta = E_m/E_s$ where $E_m =
0.6468$ and $E_s=1.7$. The shaded area is the rectangle
$[-E_s,E_s]\times [-v_s,0]$ that contains the range of the scaled
Hamiltonian $H_{s}=H/\beta$ as explained in Section 5. The maximum
imaginary part of the scaled Hamiltonian, $v_{s}=0.01$, is
consistent with our choice of the absorbing layer. The order $n$
of the Faber polynomial approximation is set by
the exponential decay of the expansion coefficients (\ref{3.8}).
In our simulations, we demand that
$|c_{k}|$ becomes less than $10^{-15}$ for $k\geq n$. The behavior
of $|c_k|$ is shown in Figure 1b 
for time steps $\Delta t_F = j\Delta t_L$ with $j=50, 200,$ and $1000$. 

The transmitted signal is collected on the ``virtual detector'' located at
$z_{d}=3.22~D_{g}$ behind the structure. The
zero-order component of the electric field,
\begin{equation}
\label{6.1}
E_{0}(z_{d},t)=\frac{1}{D_{g}}\int^{D_{g}}_{0} E(x,z_{d},t) dx\ ,
\end{equation}
is shown in Fig. 2.

The existence of a trapped mode (resonance) can easily be inferred
from the temporal evolution of the electromagnetic field. The main
transmitted pulse is clearly visible. It has a significant amplitude
and a duration about $38$ fs. After the main pulse passes the array,
it leaves behind excited quasistationary modes which loose their
energy by radiating almost monochromatic waves. By symmetry, the
same radiation of quasistationary modes 
is registered in the reflection direction by a
detector placed in front of the layer (not shown here). 
The quasistationary mode associated with polaritonic 
excitations in the ionic crystal has a wave length $\lambda
\sim D_T$ and is short-lived due to the strong absorption
of the material at the resonance (the anomalous dispersion
region). Therefore the observed lasing effect is mainly 
due to the long-lived structure resonance at $\lambda
\sim D_\rho$.
The radiation of this mode appears as exponentially damped
oscillations coming after the main signal. An exponential decay due
to a finite lifetime of the quasi-stationary mode is clearly seen.
The resonance lifetime is in the picosecond range, i.e., a thousand
times longer than the initial pulse duration. For lossless media,
the existence of the quasistationary mode(s) leads to a 100\% reflection
at the resonant frequency, as has been discussed in detail in Refs.
\cite{as2,havier}. Finally, the concept of trapped modes localized
on successive layers and interacting with each other provides a
theoretical framework for light propagation in layered structures
such as photonic crystal slabs \cite{Slabs}.

The main results of the paper are summarized in Fig. 3 and Table 1
where we show the precision of the Faber propagation scheme and
compare its numerical costs with those of the Lanczos-Arnoldi scheme.
Figure 3 presents a relative error of the time propagation, defined as $\left
| \{E_{0}(z_{d},t)-E_{ref}(z_{d},t)\}/E_{ref}(z_{d},t) \right | $,
where the reference signal $E_{ref}(z_d,t)$ is chosen to be the result
obtained by the Faber propagation scheme with $\Delta t_F = 50\Delta t_L$.
The choice is motivated by a higher precision of the Faber scheme
(thanks to its exponential convergence) and by the fact that
the factor $\max_\Gamma |\exp(-i\Delta t_F z)|$ is minimal
for the smallest $\Delta t_F$ used in our simulations.
There is no difference between the results with $\Delta_F =25\Delta t_L$
and $\Delta_F =50\Delta t_L$ (see below). 

It follows from our results that the Faber
propagation scheme has a higher accuracy than the
Lanczos-Arnoldi propagation at reduced computational costs. The error
was saturated at $10^{-10}$ value when 10 significant digits in the
calculated signal where found to coincide. The peaks correspond
mainly to the instants of time when the oscillating electric field
is close to zero. The gain in the propagation efficiency as compared
to the Lanczos-Arnoldi scheme is twofold. First, a smaller number of
actions of
$H$ on $\Psi(t)$ is needed to obtain $\Psi(t+\Delta t_F)$.
In the Faber propagation scheme,
it is given by the order of the Faber polynomial approximation of
the fundamental solution, $N_F=n$. In the case of the
Lanczos-Arnoldi scheme, the number
of actions of $H$ on $\Psi(t)$ is given
by $N_L=K\Delta t_{F}/\Delta t_{L}$ where $K$ is the dimension of the
Krylov space.  For the precision shown in Fig. 3, $K=7$. Second, as we
have already discussed in Ref.\cite{as2}, for a typical size of the mesh as
used here, computational costs of acting by $H$ on $\Psi(t)$ are
 comparable with those of
constructing an orthonormal basis
for the Krylov space (by means of the Arnoldi process) and projecting
the Hamiltonian onto the Krylov space (a $K\times K$
Hessenberg matrix for a non-Hermitian $H$). This explains
an extra factor $2.5$ in the fourth column of Table 1.
For significantly larger sizes of the mesh, in particular, for
3D simulations, the computational costs
of acting by  $H$ on $\Psi$ should prevail,
and the gain in the computation time should simply scale as $N_{L}/N_{F}$.

It is worth noting that memory requirements are lower for
the present Faber propagation owing to the short recursion relation
(associated with an elliptic contour).
Indeed, in the case of the Lanczos-Arnoldi scheme
the number of vectors to be kept in
the operational memory equals $K$.

Finally, in the Lanczos-Arnoldi propagation scheme applied to 
the above system the time step $\Delta t_L$ exceeds the 
Courant limit $\Delta t_C= \|H\|^{-1}$ at least in three times \cite{as2}.
Therefore, the Faber propagation scheme allows one to exceed the 
Courant limit at least in 3000 times, $\Delta t_F \geq 3000 \Delta t_C$
as one can see from Table 1.
\begin{table}[h]
\caption{Numerical costs and efficiency 
of the Faber polynomial propagation} 
\centering 
\begin{tabular}{cccc} 
\\[1ex]
\hline\hline 
Time step %
&$N_{F}$: number of $H\Psi$ %
&$N_{L}$: number of $H\Psi$ %
&Computation time gain \\
units of $\Delta t_{L}$%
&operations, Faber %
&operations, Lanczos %
&$\sim 2.5~N_{L}/N_{F}$
\\ [0.5ex]
\hline 
\\ [0.5ex]
25 &  170 & 175 & 2.5 \\[1ex]
50 &  290 & 350 & 3.0 \\[1ex]
100 &  525 & 700 & 3.3 \\[1ex]
200 &  980 & 1400 & 3.6 \\[1ex]
400 &  1890 & 2800 & 3.7 \\[1ex]
1000 &  4560 & 7000 & 3.8 \\[1ex]

\hline 
\end{tabular}
\label{tab:ener}
\end{table}

\section{Conclusions}
\setcounter{equation}0

We have shown that the Faber propagation scheme can successfully 
be used in electrodynamics of passive media. The scheme 
is global in time, that is, it allows one for time steps
that exceed the Courant limit in a few orders of magnitude.
As a point of fact, the propagation can actually be carried
out in a single time step if the system in question does not
have long-lived quasistationary modes (as the structure resonance
in the example 
we have considered above). 

The essential virtue of the 
scheme is the exponential convergence, which leads to 
superior accuracy as compared to other time domain methods
in passive media. The Faber propagation scheme can therefore be
used as a benchmark, when comparing various propagation methods.
If the medium is lossless and no absorber is present, the Faber
scheme coincides with the Chebyshev propagation scheme, whose
high accuracy is well known in time domain methods in computational 
quantum physics.

Another advantage of the present Faber propagation scheme is a 
relatively low memory
demand. This, however, is essentially due to an elliptic contour
which leads to a family of Faber polynomials that are generated
by a short recursion relation. 
For example, the conventional leapfrog (time differencing) 
propagation scheme 
requires to have two arrays $\Psi(t)$ and $\Psi(t-\Delta t)$
in the operational memory to compute $\Psi(t+\Delta t)$, 
while in the Faber scheme associated
with an elliptic contour, a recursive computation of the sum 
(1.3) requires storing three arrays
$\Psi_m(t+\Delta t_F)$, $\Phi_m$, and $\Phi_{m-1}$, 
where $\Psi_m$ is the series (1.3) with $k=0,1,...,m \leq n$,
and $m=1,2,..., n$ being the recurrence running index,
$\Psi_{m+1} = \Psi_m + c_{m+1}\Phi_{m+1}$. However, 
the gain of the Faber scheme in efficiency and accuracy is enormous.

It should be noted that we have not explored a further optimization
of the present Faber propagation scheme because
our main goal was to compare it with the Lanczos-Arnoldi 
propagation scheme (which was applied to the above system and 
shown to be more accurate and efficient
than a typical finite differencing (leapfrog) scheme).
For any application, the optimization should include the following.
First, the spread of the spectrum along the real axis
is essentially determined by the smallest grid spatial step.
So, depending on the accuracy demand, $E_m$ can be reduced.
Second, the absorbing layer can also be optimized to reduce
the spread $v$ of the spectrum along the imaginary axis.
In addition, one can try to estimate (e.g., by perturbation
theory) imaginary parts of eigenvalues with large real parts
(of order $E_m$). This would lead to a tighter ellipse.
Finally, the contour shape itself can also be optimized, which,
in general, requires
a better knowledge of the spectrum of the Hamiltonian. 
Thus, for a specific problem on hands, 
the Faber propagation scheme can be made even more efficient
than the simplest example presented in our work.

\vskip 1cm
\noindent
{\bf Acknowledgments}

S.V.S. thanks the LCAM of the
Univestity of Paris-Sud and, in particular, Dr. V. Sidis for the support
and warm hospitality extended to him during his stay in Orsay.

\newpage
{\bf Figure captions}

Fig. 1a.\qquad The elliptic contour used in our simulations. Results
are presented on the complex plane of the scaled energy 
$\left ( {\rm Re}~
E,~{\rm Im}~E \right )$. The shaded rectangle contains the range of
the scaled Hamiltonian $H_{s}=H/\beta$. Further details are given
in the text.\bigskip

Fig. 1b. \qquad The log-log plot of absolute values
$|c_k|$ of the expansion coefficients versus $k$ for time steps
$\Delta t_F = j\Delta t_L$
with $j=50, 200$, and $1000$ as indicated in the inset of the 
figure.

Fig. 2. \qquad  Electric field of the zero-order transmitted wave as
a function of time measured in femtoseconds. The signal is registered
by a detector placed behind the periodic layer of ionic crystal
cylinders. The grating geometry is sketched on the inset of the
figure.
\bigskip

Fig. 3.\qquad Relative error (defined in the text) for
the zero order wave transmitted through the periodic layer of ionic
crystal cylinders. Results obtained with the Faber 
propagation scheme (full symbols) and 
the Lanczos-Arnoldi scheme (open circles)
are presented as a function of time measured in femtoseconds. The
time step for the Lanczos-Arnoldi propagation is $\Delta
t_{L}=0.138$ fs. The time step for the Faber propagation scheme is
given by $\Delta t_{F}=j~\Delta t_{L}$, where the correspondence between
different symbols and the values of $j$ is indicated in the inset of
the figure.\bigskip

\end{document}